\documentclass{aastex631}

\DeclareUnicodeCharacter{2212}{-}

\begin{document}
	
	\title{JWST Exoplanetary Worlds and Elemental Survey (JEWELS) I: High-Precision Chemical Abundances of 20 FGK Planet-Hosting Stars from JWST Cycle 2}
	
	\correspondingauthor{Qinghui Sun}
	\email{qinghuisun@sjtu.edu.cn}
	
	\author[0000-0003-3281-6461]{Qinghui Sun}
	\affiliation{Tsung-Dao Lee Institute, Shanghai Jiao Tong University, Shanghai, 200240, China}
	
	\begin{abstract}
		
		We present high-precision chemical abundances for 20 FGK stars hosting planets observed in JWST Cycle 2 GO programs. Using high-resolution, high–signal-to-noise ratio spectra from the ESO and Keck archives, we perform a strict line-by-line differential analysis relative to the Sun to derive stellar parameters and abundances of 19 elements from C to Zn. The stars span effective temperatures of 4500–6500 K and metallicities from –0.57 to +0.50 dex. The sample includes hosts of both gas giants and terrestrial planets, allowing direct comparison between stellar composition and planetary properties. Several of the giant planets orbit metal-rich stars. The detailed abundance patterns show clear chemical diversity, including carbon-enhanced but mildly metal-poor stars (TOI-824, TOI-561, TOI-1130, GJ 9827) and $\alpha$-enhanced metal-poor stars (TOI-561, GJ 9827, TOI-824). These variations trace differences in protoplanetary disk composition and may influence planetary interiors and atmospheric chemistry. The planet-hosts show a range of [C/O] ratios, and the diverse [Mg/Si] ratios may suggest varied interior compositions for their rocky planets. This homogeneous stellar abundance, together with future uniform JWST planetary atmosphere measurements, provides a foundation for exploring the planet mass–metallicity relation and the connection between stellar chemistry and planetary formation pathways. These results constitute the first step in a larger survey spanning multiple JWST cycles to systematically examine how host star composition shapes exoplanetary systems.
		
	\end{abstract}
	
	\section{Introduction}
	
	The discovery and characterization of exoplanets have advanced significantly with the \textit{James Webb Space Telescope} (JWST). Since its launch in late 2021 \citep{2023PASP..135f8001G}, JWST has expanded the scope of exoplanet science, from refining orbital and bulk properties to probing atmospheric compositions with unprecedented precision. Its infrared capabilities allow studies across diverse planetary types, including ultra-hot Jupiters, temperate sub-Neptunes, and Earth-sized rocky planets, using techniques including transit spectroscopy, phase-curve observations, and direct imaging.
	
	One of the major contributiosn of JWST is the detection and characterization of atmospheric compositions over a broad spectral range (0.6–28 $\mu$m). Water (H$_2$O), sodium (Na), and potassium (K) have been the most commonly observed species, as demonstrated in pre-JWST studies (e.g., \citealt{2002ApJ...568..377C, 2008ApJ...673L..87R, 2014ApJ...791L...9M, 2015A&A...577A..62W, 2017ApJ...834...50B, 2018Natur.557..526N, 2019MNRAS.482.1485P, 2019ApJ...887L..20W}). Beyond these, JWST has detected CO$_2$ and SO$_2$ in WASP-39b \citep{2023Natur.614..649J, 2024Natur.626..979P}, CH$_4$ in WASP-80b \citep{2023Natur.623..709B}, silicate clouds in VHS1256b \citep{2023ApJ...946L...6M}, SO$_2$ and silicate clouds in WASP-107b \citep{2024Natur.625...51D}, and H$_2$O in multiple systems \citep[e.g.,][]{2024Natur.625...51D, 2023Natur.620..292C, 2025ApJ...984L..44D}. Additional detections include CO and H$_2$S in HD189733b \citep{2024Natur.632..752F}, Na and K in several planetary atmospheres \citep{2023Natur.614..659R,2023MNRAS.524..817T}, and isotopic ratios such as $^{12}$C/$^{13}$C in VHS1256b \citep{2023ApJ...957L..36G}. These measurements provide constraints on C/O ratios, atmospheric metallicities, and temperature–pressure profiles across a wide range of exoplanets \citep{2022ARA&A..60..159W}. JWST has also demonstrated its capability to probe terrestrial planet atmospheres, such as LHS475b \citep{2023NatAs...7.1317L}. Collectively, these results are reshaping our understanding of planetary atmospheres and their chemical diversity, from gas giants to potentially habitable worlds.
	
	The chemical composition of a planet’s atmosphere offers key insights into its formation and evolution (\citealt{2022ExA....53..323H}). Two widely discussed formation pathways are the core accretion and disk instability models. In the core accretion scenario (e.g., \citealt{1996Icar..124...62P}), planetesimals gradually build a solid core, which then accretes gas from the surrounding disk over several million years. Because solids can condense and migrate within the disk, the composition of the accreted gas may differ from that of the host star (\citealt{2010fee..book..157L, 2011ApJ...743L..16O}). If a giant planet forms in the outer disk and migrates inward (e.g., \citealt{2018ARA&A..56..175D, 2019ApJ...887L..20W, 2022MNRAS.509..894H}), core accretion predicts a relationship between planetary mass and atmospheric metallicity. Some observations have shown an inverse relation between planetary mass and heavy-element content, as well as correlations with stellar metallicity (e.g., \citealt{2011ApJ...736L..29M, 2013ApJ...775...80F, 2016ApJ...832...41M, 2016ApJ...831...64T}).
	
	The disk instability scenario (e.g., \citealt{1997Sci...276.1836B}) proposes that parts of the protoplanetary disk can collapse under gravity, forming massive gaseous protoplanets within a few hundred years. Unlike core accretion, this process does not predict a clear mass–metallicity trend (e.g., \citealt{2016ApJ...829..114B, 2022A&A...665L...5K}). High metallicity increases the disk’s optical depth and temperature, which can inhibit clump formation, so disk instability may be more favorable around lower-metallicity stars \citep{2002ApJ...567L.149B}. Current observations are insufficient to favor one formation pathway conclusively.
	
	Multiple chemical species detected in exoplanet atmospheres, including H$_2$O, CO, CO$_2$, SO$_2$, Na, and K, have been used to explore connections between atmospheric composition and planetary mass (e.g., \citealt{2016Natur.529...59S, 2016ApJ...826L..16H}). The chemical abundance of the host star strongly affects the composition of its planets. For example, variations in stellar oxygen abundance can influence the composition of solid material available in the protoplanetary disk, since higher O abundances generally increase the fraction of oxygen-bearing condensates such as silicates and water ice. Models of disk chemistry and planet formation show that these changes in the disk’s oxygen budget affect the water-ice content and overall makeup of planetary building blocks (e.g., \citealt{2016A&A...590A.101B, 2017MNRAS.469.4102M, 2021Sci...374..330A}). Comparing planetary and stellar abundances is therefore crucial for constraining the mass–metallicity relation. This requires homogeneous measurements of planetary atmosphere compositions (\citealt{2023ApJS..269...31E}) and consistent stellar abundance determinations.
	
	JWST data are processed through the same pipeline, allowing planetary measurements to be treated consistently. Equally important is a uniform characterization of their host stars to ensure direct comparisons. For instance, \citet{2022AJ....164...87K} analyzed chemical abundances for 17 FGK stars in JWST Cycle 1 programs. However, a complete and homogeneous set of stellar abundances for all JWST planet hosts from all cycles is still lacking. Variations in instruments and reduction methods can introduce systematic errors, complicating direct comparisons between studies. By combining homogeneous stellar abundances with uniform JWST planetary measurements across multiple cycles, it is now possible to assemble a dataset large enough to robustly examine trends such as the planet mass–metallicity relation \citep{2024AJ....167..167S} and to investigate the connection between planet formation and stellar chemical composition.
	
	We introduce the JWST Exoplanetary Worlds and Elemental Survey (JEWELS), a homogeneous program to measure high-precision stellar abundances for JWST planet-hosting stars and to connect stellar and planetary compositions. In this paper, we derive high-precision stellar abundances for host stars of exoplanets observed in JWST Cycle 2 General Observer (GO) program, focusing on FGK stars with high-resolution, high signal-to-noise ratio (S/N) public spectra. This study represents the first step in a larger survey that will eventually extend to stars in JWST Cycles 1, 3, and 4, providing a uniform dataset to examine how stellar composition shapes planetary properties and formation pathways.
	
	\section{Target Selection and Observations}
	
	Our initial sample includes all planetary systems scheduled for observation in the JWST Cycle 2 GO program. Additional selection criteria were applied to ensure that the host stars are suitable for precise stellar characterization, as described below. Table \ref{tab:jwst_hosts2} lists all planetary systems approved in the JWST Cycle 2 GO program, along with their host star effective temperatures ($T_{\rm eff}$) and the JWST instruments assigned for the observations.
	
	To support planet formation and atmospheric studies with reliable stellar parameters and abundances, we retrieved high-resolution spectra of the host stars from the ESO (\citealt{2022SPIE12186E..0DR}) and Keck (\citealt{2014SPIE.9152E..2IT}) public archives. Only spectra of sufficient quality for abundance analysis were retained. When multiple spectra of a given target from the same instrument had insufficient S/N individually, only exposures with the same resolution, instrumental setup, and closely spaced observation times were selected, so that exposure time was the only difference. The radial velocity differences among these spectra were negligible and did not affect the co-added results. The spectra were then co-added using the {\it scombine} task in IRAF\footnote{IRAF is distributed by the National Optical Astronomy Observatories, operated by the Association of Universities for Research in Astronomy Inc., under a cooperative agreement with the National Science Foundation.}, which sums the flux at each wavelength. This procedure improves the S/N while preserving intrinsic line profiles.
	
	We required a minimum S/N of 100 per \AA\ in the final co-added spectrum to achieve the precision necessary for accurate stellar parameter and abundance determinations. Targets were excluded if the combined archival spectra did not meet this S/N threshold, lacked adequate wavelength coverage, or were not obtained at sufficiently high spectral resolution. These requirements ensure that the retained spectra yield high-precision abundances and robust constraints for interpreting planet formation and atmospheric compositions. Table \ref{tab:jwst_hosts2} also reports the spectral wavelength coverage, ground-based instruments used, original program PIs, relevant comments, and the final co-added S/N values.
	
	To maintain homogeneity and precision, we further restricted the stellar hosts to those with $T_{\rm eff}$ between 4000 and 7600 K, based on values reported in the NASA Exoplanet Archive (\citealt{2025PSJ.....6..186C, nea1, nea2}). This range corresponds to late-K through late-A dwarfs, encompassing the majority of JWST Cycle 2 planet hosts while excluding cooler stars, where molecular line blending complicates abundance analysis, and hotter stars, where shallow spectral lines limit accuracy.
	
	Several systems (e.g., MWC~758 c, 51~Eri~b, KELT-7~b, TrES-4~b, and HAT-P-65~b) were excluded because their available spectra either covered too narrow a wavelength range for reliable abundance measurements, exhibited unresolved double-lined features, showed excessive rotational broadening that caused line blending, lacked publicly available data, or had already been analyzed in \citet{2024AJ....167..167S}. These cases are noted in Table \ref{tab:jwst_hosts2} and discussed in detail in the Appendix.
	
	Our final sample consists of 20 FGK-type planet-hosting stars from the JWST Cycle 2 GO program, for which we derived high-precision stellar parameters and chemical abundances. Table \ref{tab:stellar_params} lists the derived stellar atmospheric parameters (discussed in detail below), together with the planetary orbital periods, radii, and masses compiled from the literature.
	
	\section{Stellar Atmosphere Parameters}
	
	We derived stellar atmospheric parameters for the 20 planet-hosting FGK stars using a strictly differential, line-by-line analysis relative to the Sun, following the procedures described in \citet{2025ApJ...980..179S, 2025A&A...701A.107S}. Equivalent widths (EWs) for elements with atomic number $Z \leq 30$ were measured using the \texttt{splot} task in IRAF, adopting the line list of \citet{2014ApJ...791...14M}. For each instrument, a solar reference spectrum obtained from reflected sunlight (e.g., asteroid Vesta or the Moon) was analyzed in parallel, and EWs were measured for each line in the solar spectra corresponding to the different instruments.
	
	Only lines with EWs between 10 and 150 m\AA\ were retained. Lines weaker than 10 m\AA\ are too uncertain due to noise, while those stronger than 150 m\AA\ lie outside the linear regime of the curve of growth and are more affected by damping and saturation. In a few cases, all available lines of a given element were weaker than 10 m\AA; for example, the three oxygen lines in TOI-1130. In such cases, the weak lines were retained to enable an [X/H] determination, but the resulting abundances are flagged as uncertain.
	
	Stellar atmospheric parameters were determined using the \texttt{q$^2$} code (\citealt{2014A&A...572A..48R}) with MARCS model atmospheres (\citealt{2008A&A...486..951G}), based on the measured EWs of Fe lines for the star and for the Sun. The effective temperature ($T_{\rm eff}$) was obtained by requiring excitation equilibrium of Fe I lines, the surface gravity ($\log g$) by enforcing ionization equilibrium between Fe I and Fe II, and the microturbulence ($V_t$) by minimizing trends between Fe I abundances and reduced equivalent width ($\log$ EW/$\lambda$). The parameters were iterated until all equilibrium conditions were satisfied.
	
	Uncertainties were estimated by propagating line-to-line abundance scatter and accounting for parameter covariances. The complete line list and EW measurements for all elements are provided in Table \ref{tab:ew_measurements}, and the resulting $T_{\rm eff}$, $\log g$, $v_t$, [Fe/H], and their associated uncertainties are summarized in Table \ref{tab:stellar_params}.
	
	\begin{longrotatetable}
		\begin{deluxetable}{lccccccccccccc}
			\tablecaption{Stellar and planetary parameters for 20 planet-hosting F, G, K stars in JWST cycle 2 GO program \label{tab:stellar_params}}
			\tablewidth{0pt}
			\tablehead{
				\colhead{Star$^a$} &
				\colhead{$T_{\rm eff}^a$} & \colhead{$\sigma T_{\rm eff}^a$} &
				\colhead{$\log g^a$} & \colhead{$\sigma \log g^a$} &
				\colhead{[Fe/H]$^a$} & \colhead{$\sigma$[Fe/H]$^a$} &
				\colhead{$V_t^a$} & \colhead{$\sigma V_{t}^a$} &
				\colhead{Planet$^b$} & \colhead{P$^b$ (days)} & \colhead{$R_p^b$} & \colhead{$M_p^b$} \\
				\colhead{} &
				\colhead{K} & \colhead{K} &
				\colhead{dex} & \colhead{dex} &
				\colhead{dex} & \colhead{dex} &
				\colhead{km s$^{-1}$} & \colhead{km s$^{-1}$} &
				\colhead{} & \colhead{days} & \colhead{} & \colhead{}
			}
			\startdata
			WASP-121  & 6523 & 101 & 4.22 & 0.19 & 0.25 & 0.07 & 1.83 & 0.16 & WASP-121 b & 1.27 & $1.865\,R_{\rm J}$ & $1.183\,M_{\rm J}$  \\
			WASP-94 A  & 6285 &  60 & 4.38 & 0.15 & 0.40 & 0.05 & 1.47 & 0.10 & WASP-94 A b & 3.950 & $1.72\,R_{\rm J}$ & $0.452\,M_{\rm J}$ \\
			WASP-69   & 4874 & 100 & 4.33 & 0.20 & 0.22 & 0.07 & 1.80 & 0.20 & WASP-69 b & 3.868 & $\sim0.98\,R_{\rm J}$ & $\sim0.26\,M_{\rm J}$ \\
			WASP-52   & 5036 &  56 & 4.31 & 0.11 & 0.24 & 0.05 & 0.81 & 0.20 & WASP-52 b & 1.7498 & $1.27\,R_{\rm J}$ & $0.46\,M_{\rm J}$ \\
			WASP-47   & 5472 &  18 & 4.21 & 0.04 & 0.40 & 0.02 & 0.80 & 0.06 & WASP-47 e & 4.159 & $1.14\,R_{\rm J}$ & $1.14\,M_{\rm J}$ \\
			WASP-18   & 6314 &  79 & 4.25 & 0.16 & 0.07 & 0.06 & 1.64 & 0.14 & WASP-18 b & 0.941 & $1.17\,R_{\rm J}$ & $\sim10\,M_{\rm J}$ \\
			WASP-15   & 6424 &  64 & 4.29 & 0.14 & 0.09 & 0.04 & 1.26 & 0.10 & WASP-15 b & 3.7521 & $1.43\,R_{\rm J}$ & $0.54\,M_{\rm J}$ \\
			TOI-824   & 4977 &  90 & 4.52 & 0.13 & -0.20 & 0.09 & 0.97 & 0.22 & TOI-824 b & 1.393 & $2.93\,R_{\oplus}$ & $18.5\,M_{\oplus}$ \\
			TOI-561  & 5378 &  22 & 4.51 & 0.04 & -0.38 & 0.02 & 0.83 & 0.06 & TOI-561 b & 0.446 & $1.37\,R_{\oplus}$ & $\sim3.2\,M_{\oplus}$ \\
			TOI-1130  & 4504 & 100 & 4.10 & 0.14 & -0.14 & 0.10 & 0.80 & 0.24 & TOI-1130 b & 4.070 & $3.56\,R_{\oplus}$ & $19.3\,M_{\oplus}$ \\
			& &  &  &  &  &  &  &  & TOI-1130 c & 8.4 & $13.32\,R_{\oplus}$ & $325.59\,M_{\oplus}$ \\
			TOI-125   & 5204 &  25 & 4.29 & 0.06 & -0.11 & 0.02 & 0.80 & 0.14 & TOI-125 b & 4.650 & $2.73\,R_{\oplus}$ & $9.5\,M_{\oplus}$ \\
			&  &  &  &  &  &  &  &  & TOI-125 c & 9.15 & $2.76\,R_{\oplus}$ & $6.63\,M_{\oplus}$ \\
			&  &  &  &  &  &  &  &  & TOI-125 d & 19.98 & $2.93\,R_{\oplus}$ & $13.6\,M_{\oplus}$ \\
			PH-2$^{c}$ & 5734$^{+84}_{-82}$ & \nodata & 4.47$^{+0.03}_{-0.04}$ & \nodata & 0.017$^{+0.051}_{-0.044}$ & \nodata & 1.08 & \nodata & PH-2 b & 282.5 & 0.833$R_J$ & 0.274$M_J$ \\
			NGTS-2    & 6481 & 121 & 4.00 & 0.26 & -0.13 & 0.08 & 2.13 & 0.27 & NGTS-2 b & 4.512 & $1.60\,R_{\rm J}$ & $0.74\,M_{\rm J}$ \\
			LTT 9779  & 5385 &  42 & 4.38 & 0.11 & 0.24 & 0.04 & 0.88 & 0.13 & LTT 9779 b & 0.792 & $\sim4.7\,R_{\oplus}$ & $\sim29\,M_{\oplus}$ \\
			Kepler-12 & 6128 &  41 & 4.35 & 0.10 & 0.17 & 0.03 & 1.30 & 0.07 & Kepler-12 b & 4.438 & $1.70\,R_{\rm J}$ & $0.43\,M_{\rm J}$ & \\
			HD 106315 & 6577 &  98 & 4.76 & 0.17 & -0.14 & 0.06 & 2.09 & 0.25 & HD 106315 c & 9.5539 & $2.50\,R_{\oplus}$ & $\sim12\,M_{\oplus}$ \\
			HAT-P-30  & 6338 &  16 & 4.45 & 0.04 & 0.06 & 0.01 & 1.52 & 0.03 & HAT-P-30 b & 2.812 & $1.34\,R_{\rm J}$ & $0.71\,M_{\rm J}$ \\
			GJ 9827   & 4977 &  98 & 4.36 & 0.33 & -0.61 & 0.09 & 1.93 & 0.21 & GJ 9827 d & 1.21 & $1.64\,R_{\oplus}$ & $4.87\,M_{\oplus}$ \\
			GJ 504    & 6161 &  28 & 4.48 & 0.08 & 0.28 & 0.02 & 1.25 & 0.06 & GJ 504 b & 94863.5 & 1.08$R_J$ & $\sim4\,M_{\rm J}$ \\
			14 Herculis    & 5474 &  77 & 4.51 & 0.12 & 0.48 & 0.08 & 1.09 & 0.22 & 14 Herculis c & 1777.0 & $\sim$1.12$R_J$ & $\sim8.053\,M_{\rm J}$\\
			\enddata
			\tablenotetext{a}{The stellar effective temperature ($T_{\rm eff}$), surface gravity (log $g$), microturbulence ($V_t$), metallicity ([Fe/H]), and their associated uncertainties derived from high-resolution, high S/N spectra.}
			\tablenotetext{b}{References for the planetary orbital period, radius, and mass are listed below. WASP-121 b (\citealt{2016MNRAS.458.4025D}); WASP-94 A b (\citealt{2014AA...572A..49N}); WASP-69 b (\citealt{2014MNRAS.445.1114A}); WASP-52 b (\citealt{2013AA...549A.134H}); WASP-47 e (\citealt{2015ApJ...812L..18B}); WASP-18 b (\citealt{2009Natur.460.1098H}); WASP-15 b (\citealt{2009AJ....137.4834W}); TOI-824 b (\citealt{2020AJ....160..153B}); TOI-561 b (\citealt{2021AJ....161...56W}); TOI-1130 b, c (\citealt{2023AA...675A.115K}); TOI-125 b, c (\citealt{2020MNRAS.492.5399N}); PH-2 b (\citealt{2013PASP..125..889B}; \citealt{nea1,nea2}); NGTS-2 b (\citealt{2018MNRAS.481.4960R}); LTT 9779 (\citealt{2020NatAs...4.1148J}); Kepler-12 b (\citealt{2011ApJS..197....9F}); HD 106315 c (\citealt{2017AJ....153..255C}); HAT-P-30 b (\citealt{2011ApJ...735...24J}); GJ 9827 d (\citealt{2017AJ....154..266N, 2018AJ....155..148T}); GJ 504 b (\citealt{2013ApJ...774...11K}; NASA Exoplanet Archive); 14 Herculis c (\citealt{2007ApJ...654..625W}; NASA Exoplanet Archive).}
			\tablenotetext{c}{Unable to derive stellar atmospheric parameters for PH-2 via ionization balance owing to the lack of usable Fe lines; adopted literature values from \citet{2024ApJS..271...16D}.}
		\end{deluxetable}
	\end{longrotatetable}
	
	\section{Chemical Abundance Analysis}
	
	Elemental abundances were derived following the procedures described in \citet{2025ApJ...980..179S, 2025A&A...701A.107S}. We adopted a strictly line-by-line differential abundance analysis relative to the solar spectrum to compute [X/H]\footnote{[X/H] = A(X)$_{\star}$ – A(X)$_{\odot}$, where X denotes the element. A(X) = 12 + $\log(N_{\rm X}/N_{\rm H})$, and $N_{\rm X}$ and $N_{\rm H}$ are the number densities of element X and hydrogen, respectively.}. The measured EWs were analyzed with the {\it abfind} driver in MOOG to derive individual line abundances, A(X), for each element. The selected lines include those important for tracing planetary atmospheric compositions. For oxygen, we applied the NLTE corrections from \citet{2007AA...465..271R}; for all other elements, NLTE effects are expected to be small and largely cancel in the differential analysis.
	
	Final elemental abundances were obtained by averaging the [X/H] values from all lines of the same element in linear space after excluding $2\sigma$ outliers. We computed both the standard deviation ($\sigma$) and the standard deviation of the mean ($\sigma_\mu = \sigma / \sqrt{N}$). For elements measured from a single line (e.g., K), $\sigma_\mu$ was estimated from the 1$\sigma$ uncertainty in EW following \citet{2025ApJ...980..179S}, based on the plate scale and S/N of the spectrum.
	
	Table \ref{tab:abundances} reports $\sigma_\mu$, the propagated uncertainties from the stellar atmosphere parameters, and the total uncertainties obtained by adding these components in quadrature. Most elements show high-precision measurements, with typical total uncertainties ranging from 0.02 to 0.10 dex. For [Fe/H], we recommend using the uncertainties in Table \ref{tab:abundances} rather than those in Table \ref{tab:stellar_params}, as the former reports $\sigma_{\mu}$ combined with atmospheric parameter uncertainties, providing a more comprehensive estimate of the total error in the stellar [Fe/H] abundances.
	
	\begin{deluxetable*}{lcccc ccccc ccccc c}
		\tablecaption{Stellar Abundances for 20 Planet-Host F, G, K Stars in JWST cycle 2 GO program \label{tab:abundances}}
		\tablewidth{0pt}
		\tablehead{
			\colhead{Atom} & 
			\multicolumn{4}{c}{WASP-121} & \colhead{} &
			\multicolumn{4}{c}{WASP-94 A} & \colhead{} &
			\multicolumn{4}{c}{WASP-69} & ... \\
			\cline{2-5} \cline{7-10} \cline{12-15}
			& \colhead{[X/H]$^a$} & \colhead{$\sigma_\mu^a$} & \colhead{err$_{\rm atm}^a$} & \colhead{err$_{\rm comb}^a$} &&
			\colhead{[X/H]} & \colhead{$\sigma_\mu$} & \colhead{err$_{\rm atm}$} & \colhead{err$_{\rm comb}$} &&
			\colhead{[X/H]} & \colhead{$\sigma_\mu$} & \colhead{err$_{\rm atm}$} & \colhead{err$_{\rm comb}$} & ...
		}
		\startdata
		Fe (26) & 0.251 & 0.010 & 0.031 & 0.033 && 0.397 & 0.008 & 0.050 & 0.051 && 0.215 & 0.017 & 0.049 &	0.052 & ... \\
		C (6)  & 0.131 & 0.056 & 0.024 & 0.061 && 0.253 & 0.031 & 0.048 & 0.057 && 0.600 & 0.197 & 0.099	& 0.22 & ... \\
		O (8)$^b$  & 0.281 & 0.010 & 0.037 & 0.038 && 0.189 & 0.025 & 0.057 & 0.062 && 0.653 & 0.047 & 0.166 & 0.173 & ... \\
		Na (11)& 0.435 & 0.096 & 0.016 & 0.097 && 0.487 & 0.127 & 0.028 & 0.130 && 0.462 & 0.070 & 0.126 &	0.144 & ... \\
		Mg (12)& 0.206 & 0.085 & 0.027 & 0.089 && 0.301 & 0.021 & 0.032 & 0.038 && 0.171 & 0.010 & 0.069 &	0.070 & ... \\
		Al (13)	& 0.34	& 0.03	& 0.029	& 0.042	&& 0.388	& 0.029	& 0.032 &	0.043	&& 0.491	& 0.048	& 0.087	& 0.099 & ... \\
		Si (14)	& 0.295	& 0.02	& 0.016	& 0.026	&& 0.372	& 0.013	& 0.021 &	0.025	&& 0.494	& 0.042	& 0.052	& 0.067 & ... \\
		S (16)	& 0.288	& 0.072	& 0.014	& 0.073	&& 0.144	& 0.004	& 0.042 &	0.042 &&	1.207	& 0.012	& 0.097	& 0.098 & ... \\
		K (19)$^c$	& -- & -- & -- & -- && -- & -- & -- & -- && -- & -- & -- & -- & ... \\
		Ca (20)	& 0.269	& 0.028	& 0.031	& 0.042	&& 0.393 & 0.015 & 0.048 &	0.050 && 0.197	& 0.062	& 0.141	& 0.154 & ... \\
		Sc (21)	& 0.587	& 0.136 & 0.034	& 0.140	&& 0.48 & 0.032	& 0.062 &	0.070 && 0.411 & 0.097	& 0.083	& 0.128 & ... \\
		Ti (22)	& 0.418	& 0.063	& 0.041	& 0.075	&& 0.457 & 0.018 & 0.064 &	0.066 && 0.195 &0.047 & 0.067 & 0.082 & ... \\
		V (23)	& 0.368	& 0.116	& 0.059	& 0.130	&& 0.418	& 0.012	& 0.054 &	0.055	&& 0.373 & 0.07	& 0.15	& 0.166 & ... \\
		Cr (24)	& 0.27	& 0.037	& 0.035	& 0.051	&& 0.397 & 0.016 & 0.051 &	0.053	&& 0.373 & 0.089 & 0.092 & 0.128 & ... \\
		Mn (25)	& 0.447	& 0.177	& 0.033	& 0.180	&& 0.362 & 0.031 & 0.052 &	0.061	&& 0.458 & 0.158 & 0.079 & 0.177 & ... \\
		Co (27)	& 0.661	& 0.159	& 0.05	& 0.167	&& 0.389 & 0.029 & 0.047 &	0.055	&& 0.419 & 0.025 & 0.026 & 0.036 & ... \\
		Ni (28)	& 0.200	& 0.021	& 0.029	& 0.036	&& 0.417 & 0.01	& 0.041 &	0.042	&& 0.340 &	0.025	& 0.012 &	0.028 & ... \\
		Cu (29)	& 0.87	& 0.129	& 0.041	& 0.135	&& 0.39	& 0.057	& 0.036	& 0.067	&& 0.827 & 0.241 & 0.018 & 0.242& ...  \\
		Zn (30)	& 0.035	& 0.074	& 0.032	& 0.081	&& 0.286 & 0.001 & 0.035 &	0.035	&& 0.332 & 0.115 & 0.045 & 0.123 & ... \\
		\enddata
		\tablecomments{a}{Averaged abundance [X/H], standard deviation of the mean ($\sigma_{\mu}$), uncertainties propagated from the atmosphere, and co-added total uncertainty.}
		\tablecomments{b}{For Oxygen, we applied the NLTE corrections from \citet{2007AA...465..271R}, and show the NLTE-corrected [O/H] here.}
		\tablecomments{c}{For potassium (K), only a single strong line is available. Since lines with equivalent widths greater than 150 m\AA are excluded from the analysis, and in some cases the spectra do not cover the K line, the abundance A(K) is not always available.}
		\\The full table including all stars is available in machine-readable form online.
	\end{deluxetable*}
	
	As noted in previous work (e.g. \citealt{2024A&A...688A.193D}), for dwarfs cooler than 5000 K, atomic carbon lines are often too weak to be measured confidently or are strongly blended with molecular features. We observe the same behavior in our sample: for the four cool dwarfs TOI-1130, TOI-824, GJ 9827, and WASP-69, carbon abundances derived from atomic-line EWs are too high. We therefore derive carbon abundances by synthesizing the C$_2$ molecular bands at 5165 \AA\ (\citealt{2024A&A...688A.193D}) using the \textit{synth} driver in MOOG. The adopted abundances in Table \ref{tab:abundances} correspond to the best-fitting synthetic spectra, which simultaneously reproduce the C$_2$ band features and the surrounding continuum and neighboring lines.
	
	\section{Discussion}
	
	The JWST planet-host sample investigated in this study is inherently shaped by the mission’s proposal-driven target selection and therefore does not represent a statistically unbiased set of FGK dwarf stars. These systems were chosen primarily because their planets exhibit unusual or scientifically compelling characteristics, rather than because they form a representative subset of the broader exoplanet population. As a result, any apparent abundance patterns in this section should be interpreted strictly within the context of this curated sample and not extrapolated to the general population of planet-hosting stars. The figures presented here are intended to provide a global view of the diversity of stellar abundance behavior within the JWST targets, rather than to identify universal or population-level astrophysical trends.
	
\begin{figure*}
		\centering
		\includegraphics[width=1.0\textwidth]{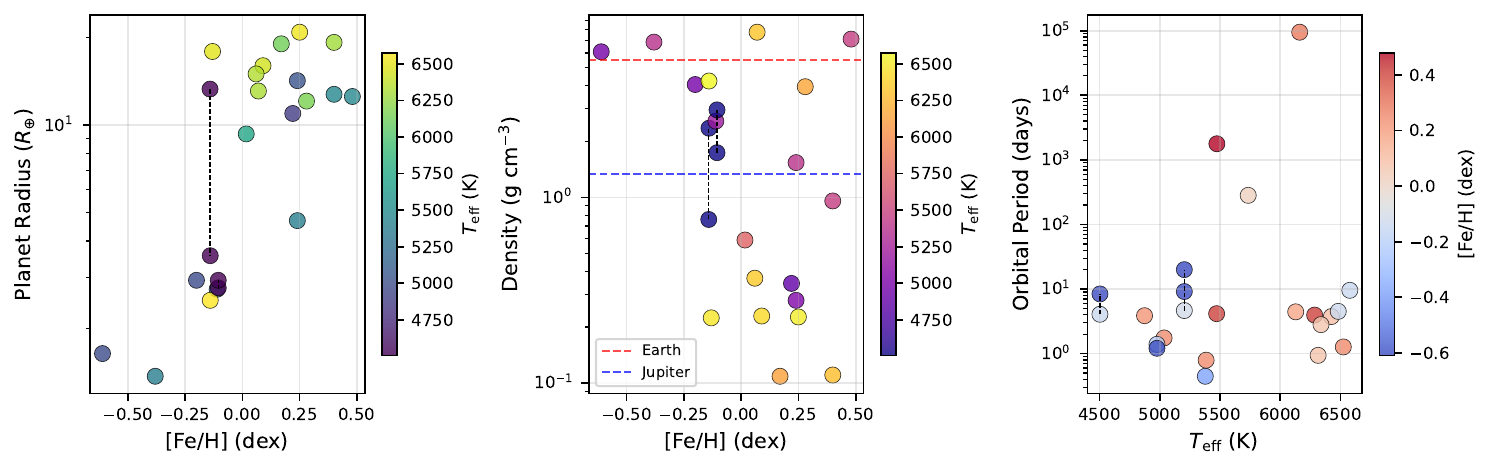}
		\caption{Relations between stellar and planetary properties within the JWST Cycle 2 GO sample. (a) Planetary radius versus stellar metallicity. (b) Planetary density versus metallicity. (c) Orbital period versus stellar $T_{\rm eff}$. Colors indicate $T_{\rm eff}$ (a,b) and [Fe/H] (c). For TOI-1130 and TOI-125, which host multiple planets, all planets are shown and connected with black dashed lines.}
		\label{fig1}
	\end{figure*}

	The planet-hosting FGK stars in the JWST Cycle 2 GO sample exhibit a wide range of chemical compositions, with [Fe/H] values spanning from $-0.57$ to $+0.50$ dex. \citet{2005ApJ...622.1102F} found that stars with extrasolar planets are more likely to form in higher-metallicity environments. Within this sample, we likewise find that many giant-planet hosts have above-solar metallicities. Figure \ref{fig1} plots several relationships between stellar and planetary properties as they appear in this curated sample. Panel (a) shows planetary radius as a function of stellar metallicity, colored by $T_{\rm eff}$. While some gas giants with large radii appear around metal-rich stars such as WASP-94 A, WASP-47, and 14 Herculis, removing the two most metal-poor stars substantially weakens any apparent trend. Panel (b) presents planetary density versus metallicity, where a few low-density (inflated) planets orbit more metal-rich and hotter stars. Panel (c) shows orbital period versus stellar $T_{\rm eff}$, colored by [Fe/H]; here, several long-period giants (e.g., GJ 504 b) are metal-rich and orbit relatively hotter stars.
	
	Beyond overall metallicity, detailed abundance patterns reveal additional diversity. The sample includes stars with near-solar compositions and others showing distinct chemical peculiarities, such as carbon-enhanced but slightly metal-poor stars (TOI-824, TOI-561, TOI-1130, GJ 9827), volatile-rich stars (WASP-69, WASP-52), and old, $\alpha$-enhanced, metal-poor stars (TOI-561, GJ 9827). The overall $\alpha$-element abundance is computed as the mean of the Mg, Si, Ca, and Ti abundances, where measurements for these elements are available. These abundance differences trace the composition of the natal protoplanetary disks and can influence planet formation outcomes. For example, the stellar [C/O] affects the atmospheric chemistry of gas giants, while refractory element ratios such as [Mg/Si] shape the interior composition of terrestrial planets. Together, these stellar chemical signatures provide essential context for interpreting the structure, bulk properties, and atmospheres of the associated planetary systems.
	
	\begin{figure*}
		\centering
		\includegraphics[width=1.0\textwidth]{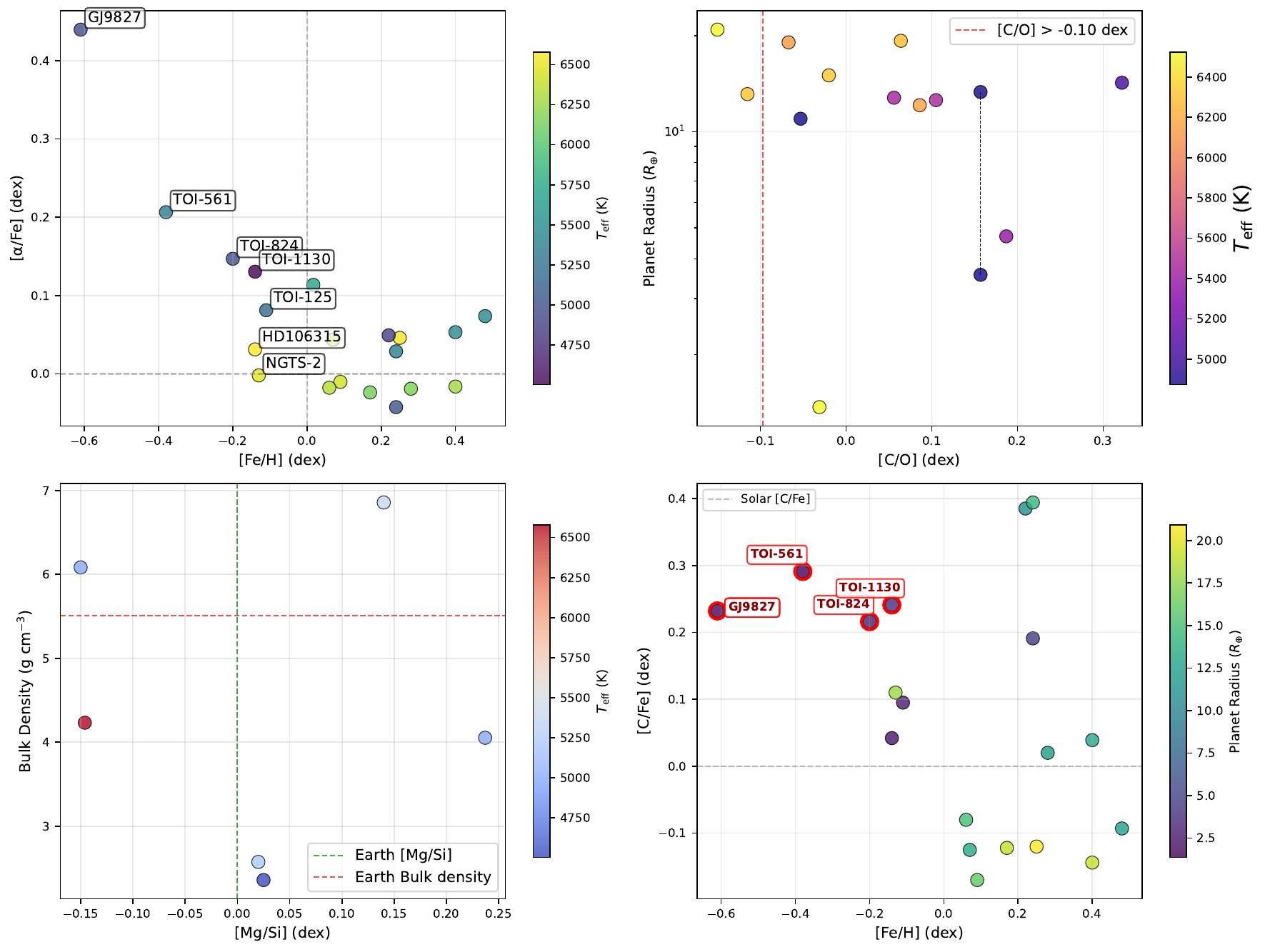}
		\caption{Relation between stellar chemistry and planetary systems within the JWST Cycle 2 GO sample. (a) $\alpha$-element abundance ratio versus metallicity, tracing Galactic chemical evolution. (b) Planetary radius versus stellar [C/O], where high [C/O] may indicate carbon-rich atmospheres. Both planets in the TOI-1130 system are shown and connected by a black dashed line. TOI-125 lacks an [O/H] measurement and is not shown. (c) Bulk density of rocky planets versus [Mg/Si]. The planets TOI-1130 b and TOI-125 b are included in this panel. (d) Stellar carbon enhancement ([C/Fe]) versus [Fe/H], highlighting carbon-rich while slightly metal-poor stars.}
		\label{fig2}
	\end{figure*}
	
	Several stars in the sample show carbon enrichment (Figure~\ref{fig2}d). GJ 9827, TOI-561, TOI-824, and TOI-1130 have [C/Fe] values of 0.23, 0.291, 0.216, and 0.241 dex, respectively, despite having slightly subsolar metallicities ([Fe/H] = -0.612, -0.379, -0.196, and -0.141 dex). One possible explanation is mass transfer from an evolved asymptotic giant branch (AGB) companion (e.g., \citealt{2005ApJ...625..825L, 2010A&A...509A..93M}). Enrichment from early supernovae is unlikely because these stars do not reach the very low metallicities of [Fe/H] $< -1.0$ dex. Apart from carbon-enriched metal-poor stars, WASP-69 also exhibits elevated C, O, and S abundances, indicative of a volatile-rich formation environment.
	
	To examine whether mass transfer is feasible for GJ 9827, TOI-561, TOI-824, and TOI-1130, we first check for evidence of companions. As JWST targets, these stars have extensive RV follow-up, which should reveal any nearby massive companion. The literature shows no sign of a secondary object around TOI-824, TOI-561, or GJ-9827. For TOI-1130, \citet{2023A&A...675A.115K} report a linear RV trend that may indicate an outer massive companion, though this has not yet been confirmed. We next examine the abundances of s-process elements (Sr, Y, Ba), which can indicate past AGB mass transfer. We perform spectrum synthesis using the {\it synth} driver in MOOG for Sr, Y, and Ba, and apply the same procedure to a solar spectrum taken with the same instrument. Apart from TOI-561, the rest three stars show strong Sr absorption lines, with EWs more than twice those of the Sun, indicating significant Sr enrichment. The Y and Ba lines are also detected though in general weaker than the solar lines. Table \ref{tab:s-process} lists the synthesized abundances for each line and the mean abundances calculated in linear space for GJ 9827, TOI-824, and TOI-1130. The uncertainties are estimated in the same way as in previous sections, coadding the propagated atmospheric-parameter errors and $\sigma_{\mu}$ in quadrature. For GJ 9827, TOI-824, and TOI-1130, Sr is above solar and strongly enhanced relative to iron, with [Sr/Fe] values of 1.49, 1.07, and 0.91 dex. In contrast, Y and Ba are below solar in all cases, but their absolute abundances are consistent with the stars’ slightly subsolar metallicities. Although Sr is enhanced, the lack of corresponding Y and Ba enhancement may not suggest mass transfer from an AGB companion.
	
	\begin{deluxetable*}{ccccccccccccc}
		\tablecaption{Sr, Y, and Ba abundance for three carbon-enhanced, metal poor stars \label{tab:s-process}}
		\tablewidth{0pt}
		\tablehead{
			\colhead{Wavelength$^a$} &  \colhead{Ion$^a$} &  \colhead{EP$^a$} &  \colhead{log({\it gf})$^a$} & $C_6^a$ & \colhead{HARPS$^b$} & 
			\colhead{ESPRESSO$^b$} & \multicolumn{2}{c}{GJ 9827$^c$} &  \multicolumn{2}{c}{TOI-824$^c$} &  \multicolumn{2}{c}{TOI-1130$^c$} \\
			\colhead{} &  \colhead{} &  \colhead{} &  \colhead{} & & \colhead{A(X)} & \colhead{A(X)} &  \colhead{A(X)} &  \colhead{[X/H]} & \colhead{A(X)} & \colhead{[X/H]} & \colhead{A(X)} & \colhead{[X/H]} \\
			\colhead{\AA} &  \colhead{} &  \colhead{eV} &  \colhead{} & & \colhead{dex} & \colhead{dex} &  \multicolumn{2}{c}{dex} & \multicolumn{2}{c}{dex} & \multicolumn{2}{c}{dex}
		}
		\startdata
		4607.338 & 38.0 & 0     & 0.283  & 6.55E$-$32 & 2.67 & 2.67 & 3.55 & 0.88 & 3.54 & 0.87 & 3.44 & 0.77 \\
		4883.685 & 39.1 & 1.0841 & 0.07   & 2.80E$+$00 & 1.91 & 1.96 & 1.64 & $-0.27$ & 1.98 & 0.07 & 1.86 & $-0.10$ \\
		5087.420 & 39.1 & 1.0841 & $-0.17$ & 2.80E$+$00 & 2.08 & 2.11 & 1.24 & $-0.84$ & 1.58 & $-0.50$ & 1.61 & $-0.50$ \\
		5853.670 & 56.1 & 0.604  & $-0.91$ & 5.30E$-$32 & 1.51 & 1.50 & 0.71 & $-0.80$ & 1.30 & $-0.21$ & 1.30 & $-0.20$ \\
		6141.710 & 56.1 & 0.704  & $-0.08$ & 5.30E$-$32 & 1.80 & 1.80 & 1.11 & $-0.69$ & 1.85 & 0.05 & 1.83 & 0.03 \\
		6496.900 & 56.1 & 0.604  & $-0.38$ & 5.30E$-$32 & 1.63 & 1.76 & 0.81 & $-0.82$ & 1.58 & $-0.05$ & 1.53 & $-0.23$ \\
		&  &   &   &  &  \multicolumn{2}{c}{[Sr/H]$^d$ = }  & \multicolumn{2}{c}{0.88 $\pm$ 0.14} & \multicolumn{2}{c}{0.87 $\pm$ 0.09} & \multicolumn{2}{c}{0.77 $\pm$ 0.11} \\
		&  &   &   &  &  \multicolumn{2}{c}{[Y/H]$^d$ = }  & \multicolumn{2}{c}{-0.47 $\pm$ 0.22} & \multicolumn{2}{c}{-0.13 $\pm$ 0.22} & \multicolumn{2}{c}{-0.26 $\pm$ 0.15} \\
		&  &   &   &  &  \multicolumn{2}{c}{[Ba/H]$^d$ = }  & \multicolumn{2}{c}{-0.77 $\pm$ 0.04} & \multicolumn{2}{c}{-0.06 $\pm$ 0.07} & \multicolumn{2}{c}{-0.12 $\pm$ 0.08} \\
		\enddata
		\tablecomments{a. Wavelength in \AA. Ion identifier: integer represents atomic number, fractional part indicates ionization state (e.g., 26.0 = Fe I, 26.1 = Fe II). Excitation potential, log (gf) value, and damping constant of the line. \\
			b. Solar A(X) values synthesized from the HARPS or ESPRESSO solar spectra. \\
			c. Synthesized A(X) values for GJ 9827, TOI-824, and TOI-1130. The [X/H] ratios are computed as A(X)$_{\star}$ - A(X)$_{\odot}$. GJ 9827 and TOI-824 use HARPS solar abundances, and TOI-1130 uses ESPRESSO solar abundances. \\
			d. Final mean [Sr/H], [Y/H], and [Ba/H] values and their uncertainties. The uncertainties combine the propagated atmospheric-parameter errors and $\sigma_{\mu}$, added in quadrature.}
	\end{deluxetable*}
	
	The $\alpha$-element ratios further reveal the Galactic origins of these stars. Elements such as Mg, Si, Ti, and Ca, primarily produced in core-collapse supernovae, trace early star formation timescales. Metal-poor stars including GJ 9827, TOI-561, and TOI-824 show high [$\alpha$/Fe] (Figure~\ref{fig2}a), consistent with rapid formation before substantial iron enrichment from Type Ia supernovae. These stars likely belong to the thick-disk population. In contrast, metal-rich hosts such as WASP-47, WASP-94 A, and 14 Herculis exhibit near-solar or slightly subsolar [$\alpha$/Fe], typical of thin-disk stars formed over longer timescales.
	
	In Figure \ref{fig2}b, planet radius is plotted against [C/O]. The [C/O] values vary across systems; in particular, the multi-planet system TOI-1130 exhibits high [C/O]. The [Mg/Si] ratio also provides insight into the composition of rocky planets. As shown in Figure \ref{fig2}c, stars in the sample span a range of [Mg/Si]. This ratio controls the dominant mineralogy of terrestrial mantles \citep{2010ApJ...715.1050B, 2010ApJ...725.2349D}: low [Mg/Si] values favor feldspar-rich silicates, while high [Mg/Si] values produce olivine- and pyroxene-dominated interiors. Variations in this ratio suggest that rocky planets around different hosts may have diverse interior structures, mantle convection patterns, and volcanic activity, resulting in a wider range of geological and thermal behaviors than found in the Solar System.
	
	In summary, this study presents a homogeneous set of high-precision stellar abundances for planet-hosting stars observed in the JWST Cycle 2 General Observer (GO) program, spanning a range of stellar types (FGK) and Galactic populations (thin and thick disk). This dataset complements JWST measurements of exoplanet atmospheric compositions and enables a more robust interpretation of planetary chemical properties. As part of an ongoing survey, future work will extend these abundance measurements to stars observed in JWST Cycles 1, 3, and 4, establishing a complete and uniform dataset to investigate the planetary mass–metallicity relation and to test the relative roles of core accretion and disk instability in planet formation.
	
	\section*{acknowledgements}
	
	We thank Elisa Delgado Mena for helpful discussions regarding carbon abundances. This work is supported by the National Key R\&D Program of China under Grant No. 2024YFA1611801, the Science and Technology Commission of Shanghai Municipality under Grant No. 25ZR1402244, and the Shanghai Jiao Tong University Funds Program No. AF4260012. This work is also supported in part by Office of Science and Technology, Shanghai Municipal Government (grant Nos. 24DX1400100, ZJ2023-ZD-001). This work is based, in part, on observations made with the NASA/ESA/CSA James Webb Space Telescope.
	
	Some of the data presented herein were obtained at Keck Observatory, which is a private 501(c)3 non-profit organization operated as a scientific partnership among the California Institute of Technology, the University of California, and the National Aeronautics and Space Administration. The Observatory was made possible by the generous financial support of the W. M. Keck Foundation.This research has made use of the Keck Observatory Archive (KOA), which is operated by the W. M. Keck Observatory and the NASA Exoplanet Science Institute (NExScI), under contract with the National Aeronautics and Space Administration.
	
	This research is also based on data obtained from the ESO Science Archive Facility under request 080.A-9021(A), 192.C-0224(C), 093.A-9008(A), 099.C-0491(A), 0102.D-0789(D), 198.C-0169(A), 084.C-1039(A), 112.25T7.001, 099.C-0303(A), 1102.C-0923(A), 111.24HZ.001, 0103.C-0874(A), 112.25T4.001, 105.20KD.002, 082.C-0040(E), 106.21BV.009, 0103.C-0422(A), 0102.C-0493(B), 097.C-0863(A), 094.A-9010(A), 095.D-0383(A), 106.21QM.002, 0102.D-0185(A), 089.D-0202(A).
	
	\bibliography{sun25_JWST}{}
	\bibliographystyle{aasjournal}
	
	\appendix
	
	Table \ref{tab:jwst_hosts2} lists the planets included in the JWST Cycle 2 General Observer (GO) program. For each target, we provide the host star effective temperature ($T_{\rm eff}$) from NASA Exoplanet Archive, the archival source and wavelength coverage of available spectra, the principal investigator (PI), and comments on data quality or special considerations (e.g., co-added spectra, insufficient S/N, or absence of public data). Exposure times or S/N are reported where available, together with the planned JWST instrument modes. Systems are grouped by spectral type, with FGK stars listed first, followed by A stars and late-K/M dwarfs. In this paper, we focus on FGK hosts with spectra of S/N $>$ 100 and full optical coverage. Abundance measurements for M dwarfs will be presented in future work.
	
	Several individual cases are worth noting. The spectrum of MWC 758 covers too narrow a wavelength range to allow reliable abundance measurements. The 51 Eri spectrum exhibits blended double lines that cannot be separated for individual component analyses. KELT-7 is a very rapid rotator, and its spectrum is heavily blended and thus unsuitable for standard FGK abundance techniques, so it is excluded from this study. No public spectra are available for TrES-4 b or HAT-P-65 b. Other planet hosts were previously analyzed in \citet{2024AJ....167..167S}, as noted in the table, and are not re-measured here. In total, we derive high-precision stellar parameters and chemical abundances for 20 FGK stars in the JWST Cycle 2 GO sample.
	
	For the strict differential abundance analysis, each star is compared to a solar spectrum obtained with the same instrument. Depending on public availability, we use either a daytime sky solar spectrum or an asteroid (e.g., Vesta) reflecting sunlight. The ESPRESSO Vesta spectrum was obtained from the ESO Archive (Program 0102.D-0185(A)). The Keck/HIRES Vesta spectrum, taken by M. Asplund on 2015 November 1, was retrieved from the Keck Archive. The high S/N HARPS day-sky solar spectrum provided by T. H. Dall (used in \citealt{2006A&A...454..341D}) was downloaded from the ESO website.\footnote{\url{https://www.eso.org/sci/facilities/lasilla/instruments/harps/inst/monitoring/sun.html}} The FEROS Vesta spectrum (taken on date 2012 December 22) was obtained from the ESO Archive, and the UVES Vesta spectrum was retrieved from the ESO Archive under Program 089.D-0202(A). All of the solar reference spectra have S/N greater than 200.
	
	Table \ref{tab:ew_measurements} provides the equivalent width (EW) measurements used in the stellar parameter and abundance analysis. For each line, we list the wavelength, ion, excitation potential, $\log(gf)$, and adopted damping constant. EW measurements from solar spectra obtained with VLT/ESPRESSO, KECK/HIRES, ESO-3.6m/HARPS, ESO-2.2m/FEROS, and VLT/UVES are included for cross-comparison, followed by those for the stellar targets. Note that for four cool dwarfs with $T_{\rm eff}$ between 4500 and 5000 K—TOI-1130, TOI-824, GJ 9827, and WASP-69—the final carbon abundances are derived by fitting the $C_2$ molecular band at 6165 \AA. EWs of the corresponding atomic C lines are still listed in the table; however, these lines are either too weak or affected by blending and are therefore provided for reference only. A subset of the table is shown here; the complete set of line measurements is available in machine-readable format.
	
	\begin{longrotatetable}
		\begin{deluxetable}{cccp{3cm}cp{3cm}cp{3cm}}
			\tablecaption{JWST Cycle 2 GO program targets and host-star metadata. \label{tab:jwst_hosts2}}
			\renewcommand\thetable{A1}
			\small
			\tablehead{
				\colhead{Planet$^a$} & \colhead{$T_{\rm eff}^b$ (K)} & \colhead{Archive$^c$} & \colhead{Wavelengtht$^c$ (\AA)} &
				\colhead{PIt$^c$} & \colhead{Note$^c$} & \colhead{Exp./S/Nt$^c$} & \colhead{JWST instrument(s)t$^a$}
			}
			\startdata
			MWC 758 c & 7580 & UVES & 3732--5000 & Martin-Zaidi, C. & spectral wavelength range too narrow & 490 & NIRCam Coronagraph \\
			51 Eri b & 7422 & FEROS & 3527--9216 & Weise, Patrick & -- & 409 & NIRSpec G395H \\
			KELT-7 b & 6789 & HIRES & 2950--7590; 5090--9630 & Alam & -- & 300 & NIRSpec G395H \\
			WASP-121 b & 6776 & ESPRESSO & 3772--7900 & Hoeijmakers, Jens & already in Cycle 1 & 465 & MIRI LRS \\
			NGTS-2 b & 6478 & HARPS & 3782--6913 & Bouchy, Francois & co-add 7 exposures & -- & NIRSpec G395H \\
			WASP-18 b & 6400 & ESPRESSO & 3772--7900 & Pino, Lorenzo & -- & 1366 & NIRISS SOSS \\
			HAT-P-30 b & 6338 & ESPRESSO & 3772--7900 & Cegla, H. M. & -- & $>$300 & NIRSpec G395H \\
			WASP-15 b & 6300 & HARPS & 3782--6913 & Cameron, A. & co-add 5 exposures & $>$100 & NIRSpec G395H \\
			TrES-4 b & 6295 & -- & -- & -- & public spectra not found & -- & NIRSpec G395H \\
			HD 106315 c & 6260 & HARPS & 3782--6913 & Santerne, Alexandre & -- & 160 & MIRI LRS \\
			WASP-94 A b & 6194 & UVES & 3064--3917; 4726--10436 & Ramirez, Ivan; Mortier, A. & -- & $>$100 & NIRSpec G395H \\
			GJ 504 b & 6000 & FEROS & 3527--9216 & Kopytova, Taisiya & -- & 127 & MIRI MRS \\
			HAT-P-1 b & 5975 & -- & -- & -- & published in Sun24 & -- & MIRI LRS \\
			Kepler-12 b & 5947 & HIRES & 3360--8100 & Hillenbrand & check the spectra & 1000s & NIRSpec PRISM \\
			HAT-P-65 b & 5872 & -- & -- & -- & public spectra not found & -- & NIRSpec PRISM \\
			PH-2 b & 5734 & HIRES & 3360--8100 & Muirhead & -- & 839s & NIRSpec PRISM \\
			WASP-47 e & 5552 & ESPRESSO & 3772--7900 & Bayliss, Daniel & co-add multiple spectra & 230 & NIRSpec G395H \\
			WASP-96 b & 5500 & -- & -- & -- & published in Sun24, but S/N $<$100 & -- & NIRSpec G395H \\
			WASP-39 b & 5485 & -- & -- & -- & published in Sun24, but S/N $<$100 & 1366 & NIRISS SOSS; NIRCam F322W2 grism; NIRSpec G395H; NIRSpec PRISM \\
			LTT 9779 b & 5445 & ESPRESSO & 3772--7900 & Vaughan, Sophia & -- & 300 & NIRSpec G395H \\
			TOI-561 b & 5326 & ESPRESSO & 3772--7900 & Soares, Bárbara & -- & 174 & NIRSpec G395H \\
			TOI-125 b, c & 5320 & HARPS & 3782--6913 & Gandolfi, Davide & co-add 6 exposures & 150 & NIRSpec G395H \\
			14 Herculis c & 5129 & HIRES & 3360--8100 & Konacki & -- & 228s & NIRCam Coronagraph \\
			WASP-52 b & 5000 & ESPRESSO & 3772--7900 & Chen, Guo & -- & 160 & NIRSpec G395H \\
			HAT-P-11 b & 4780 & -- & -- & -- & published in Sun24, but S/N $<$100 & -- & MIRI LRS; NIRSpec G395H \\
			WASP-69 b & 4715 & UVES/FEROS & 3732--5000; 5655--9464; 3528--9218 & Salz, Michael & -- & 300; 500 & NIRSpec G395H; MIRI LRS \\
			TOI-824 b & 4600 & HARPS & 3782--6913 & Nielsen, Louise & -- & 83 & NIRSpec G395H \\
			WASP-43 b & 4400 & -- & -- & -- & published in Sun24, but S/N $<$100 & -- & MIRI LRS \\
			TOI-1130 b, c & 4360 & ESPRESSO & 3772--7900 & Knudstrup, Emil & -- & 246 & NIRISS SOSS; NIRSpec G395H \\
			GJ 9827 d & 4255 & HARPS & 3782--6913 & Gandolfi, Davide & co-add 24 exposures & 270 & NIRISS SOSS; NIRSpec G395H \\
			\hline
			\multicolumn{8}{c}{A dwarf; late-K/M dwarfs} \\
			\hline
			HIP 65426 b & 8840 & -- & -- & -- & -- & -- & NIRCam Coronagraph; MIRI Coronagraph; NIRISS AMI \\
			TOI-3757 b & 3913 & -- & -- & -- & -- & -- & NIRSpec PRISM \\
			TOI-1899 b & 3841 & -- & -- & -- & -- & -- & NIRSpec PRISM \\
			K2-22 b & 3830 & -- & -- & -- & -- & -- & MIRI LRS \\
			HD 260655 b & 3803 & -- & -- & -- & -- & -- & MIRI F1500W \\
			HATS-75 b & 3790 & -- & -- & -- & -- & -- & NIRSpec PRISM \\
			HATS-6 b & 3770 & -- & -- & -- & -- & -- & NIRSpec PRISM \\
			TOI-3714 b & 3652 & -- & -- & -- & -- & -- & NIRSpec PRISM \\
			TOI-5293 A b & 3586 & -- & -- & -- & -- & -- & NIRSpec PRISM \\
			TOI-1231 b & 3562 & -- & -- & -- & -- & -- & NIRISS SOSS; NIRSpec G395H; MIRI LRS \\
			GJ 3090 b & 3556 & -- & -- & -- & -- & -- & NIRISS SOSS; NIRSpec G395H \\
			TOI-1685 b & 3519 & -- & -- & -- & -- & -- & NIRSpec G395H \\
			TOI-270 d & 3506 & -- & -- & -- & -- & -- & NIRISS SOSS; MIRI LRS; NIRSpec G395H \\
			L-231-32 b & 3506 & -- & -- & -- & -- & -- & MIRI F1500W \\
			GJ 357 b & 3505 & -- & -- & -- & -- & -- & MIRI F1500W \\
			TOI-1468 c & 3496 & -- & -- & -- & -- & -- & NIRISS SOSS; NIRSpec G395H; MIRI LRS \\
			TOI-1468 b & 3496 & -- & -- & -- & -- & -- & MIRI F1500W \\
			TOI-3984 A b & 3476 & -- & -- & -- & -- & -- & NIRSpec PRISM \\
			TOI-5205 b & 3430 & -- & -- & -- & -- & -- & NIRSpec PRISM \\
			L 98-59 b,c,d & 3415 & -- & -- & -- & -- & -- & NIRSpec G395H; NIRISS SOSS; MIRI LRS; MIRI F1500W \\
			COCONUTS-2 b & 3406 & -- & -- & -- & -- & -- & NIRSpec PRISM; NIRSpec G395H; MIRI LRS \\
			TOI-3235 b & 3389 & -- & -- & -- & -- & -- & NIRSpec PRISM \\
			LHS 1478 b & 3381 & -- & -- & -- & -- & -- & MIRI F1500W \\
			GJ 3473 b & 3347 & -- & -- & -- & -- & -- & MIRI F1500W \\
			TOI-2445 b & 3333 & -- & -- & -- & -- & -- & NIRSpec PRISM \\
			LTT 3780 b,c & 3331 & -- & -- & -- & -- & -- & NIRSpec G395H; NIRISS SOSS; MIRI LRS; MIRI F1500W \\
			LHS 1140 b & 3216 & -- & -- & -- & -- & -- & MIRI F1500W \\
			LHS 3844 b & 3036 & -- & -- & -- & -- & -- & NIRSpec G395H \\
			TRAPPIST-1 b,c & 2566 & -- & -- & -- & -- & -- & MIRI F1500W \\
			2M1207 b & 1600 & -- & -- & -- & -- & -- & NIRSpec IFU \\
			VHS 1256 b & 1190 & -- & -- & -- & -- & -- & NIRSpec IFU; MIRI MRS; NIRCam Imaging \\
			\enddata
			\tablecomments{
				a. Planet proposed for observation in the JWST Cycle 2 GO program, along with the JWST instrument assigned for the observation.\\
				b. Stellar effective temperature ($T_{\rm eff}$) taken from the \citet{nea1,nea2}.\\
				c. Ground-based stellar spectra information: adopted instrument, wavelength coverage, original PI of the public spectra, relevant notes, and total exposure time (for HIRES) or signal-to-noise ratio (Exp./S/N).
			}
		\end{deluxetable}
	\end{longrotatetable}
	
	\begin{longrotatetable}
		\begin{deluxetable}{ccccccccccccccccc}
			\centering
			\tablecaption{Equivalent Width Measurements for Solar and Stellar Spectra\label{tab:ew_measurements}}
			\renewcommand\thetable{A2}
			\tablewidth{0pt}
			\tablehead{
				\colhead{$\lambda^a$} & \colhead{Ion$^a$} & \colhead{$EP^a$} & \colhead{$\log(gf)^a$} & \colhead{$C_6^a$} & 
				\colhead{ESPRESSO$^b$} & \colhead{HIRES$^b$} & \colhead{HARPS$^b$} & \colhead{FEROS$^b$} & \colhead{UVES$^b$} &
				\colhead{WASP-121$^c$} & \colhead{WASP-94A$^c$} & \colhead{...$^c$} \\
				\colhead{(\AA)} & \colhead{} & \colhead{(eV)} & \colhead{} & \colhead{} & 
				\colhead{(m\AA)} & \colhead{(m\AA)} & \colhead{(m\AA)} & \colhead{(m\AA)} & \colhead{(m\AA)} &
				\colhead{(m\AA)} & \colhead{(m\AA)} & \colhead{...}
			}
			\startdata
			5044.211 & 26.0 & 2.8512 & -2.058 & 2.71E-31 & 73.3 & 73.3 & 72.6 & 74.2 & 73.0 & 64.8 & 81.1 & ... \\
			5054.642 & 26.0 & 3.6400 & -1.921 & 4.68E-32 & 39.3 & 40.5 & 40.3 & 42.0 & 39.5 & 28.7 & 45.7 & ... \\
			5127.359 & 26.0 & 0.9150 & -3.307 & 1.84E-32 & 96.5 & 96.0 & 95.9 & 97.0 & 96.1 & 89.8 & 101.0 & ... \\
			... & ... & ... & ... \\
			\enddata
			\tablenotetext{a}{Wavelength in \AA. Ion identifier: integer represents atomic number, fractional part indicates ionization state (e.g., 26.0 = Fe I, 26.1 = Fe II). Excitation potential, log (gf) value, and damping constant of the line.}
			\tablenotetext{b}{Equivalent widths (m\AA) of the solar spectrum obtained with different instruments.}
			\tablenotetext{c}{Equivalent widths (m\AA) of the object spectrum.}
			\tablecomments{This table is available in its entirety in machine-readable form.}
		\end{deluxetable}
	\end{longrotatetable}
	
\end{document}